\documentclass{aa}
\usepackage{subfig}
\usepackage[section]{placeins}
\usepackage{float}
\usepackage{bm}
\usepackage{siunitx}
\bibpunct{(}{)}{;}{a}{}{,} 
\def\Msun{\ensuremath{\rm M_\odot}}

\def\({\left(}
\def\r){\right)}

\def\Nbh{N_\mathrm{bh}}
\def\Rbh{\sigma_\mathrm{bh}}

\usepackage{todonotes}

\usepackage{threeparttable}
\usepackage{multirow}
\usepackage{soul}

\begin{document}
\title{The Abundance of Clustered Primordial Black Holes from Quasar Microlensing}
\author{Sven ~Heydenreich\inst{1,2,3}
\and Evencio ~Mediavilla\inst{2,4}
\and Jorge ~Jim\' enez-Vicente\inst{5,6}
\and Héctor Vives-Arias \inst{7}
\and Jose A. Muñoz \inst{8,9}}
\institute{ Argelander Institute for Astronomy, University of Bonn, Auf dem H\" ugel 71, 53121 Bonn, Germany
\and Instituto de Astrof\' isica de Canarias, E-38200 La Laguna, Santa Cruz de Tenerife, Spain
\and Department of Astronomy and Astrophysics, University of California, Santa Cruz, 1156 High Street, Santa Cruz, CA 95064 USA
\and Departamento de Astrof\' isica, Universidad de La Laguna, E-38200 La Laguna, Santa Cruz
de Tenerife, Spain
\and Departamento de F\' isica Te\' orica y del Cosmos, Universidad de Granada, Campus de
Fuentenueva, E-18071 Granada, Spain
\and Instituto Carlos I de F\' isica Te\' orica y Computacional, Universidad de Granada, E-18071 Granada, Spain
\and Centro de Estudios de Física del Cosmos de Aragón (CEFCA), Plaza San Juan, 1, E-44001 Teruel, Spain
\and Departamento de Astronomía y Astrofísica, Universidad de Valencia, E-46100 Burjassot, Valencia, Spain
\and Observatorio Astronómico, Universidad de Valencia, E-46980 Paterna, Valencia, Spain
}
\abstract{While elementary particles are the favored candidate for the elusive dark matter, primordial black holes (PBHs) have also been considered to fill that role. Gravitational microlensing is a very well-suited tool to detect and measure the abundance of compact objects in galaxies. Previous studies based on quasar microlensing exclude a significant presence of substellar to intermediate-mass BHs ($\lesssim 100M_\odot$). However, these studies were based on a spatially uniform distribution of BHs while, according to current theories of PBHs formation, they are expected to appear in clusters. We study the impact of clustering in microlensing flux magnification finding that at large scales clusters act like giant pseudo-particles, strongly affecting the emission coming from the Broad Line Region, which can no longer be used to define the zero microlensing baseline. As an alternative, we set this baseline from the intrinsic magnification ratios of quasar images predicted by macro lens models and compare them with the observed flux ratios in emission lines, infrared (IR), and radio. The (magnitude) differences are the flux-ratio anomalies attributable to microlensing, which we estimate for 35 image pairs corresponding to 12 lens systems. A Bayesian analysis indicates that the observed anomalies are incompatible with the existence of a significant population of clustered PBHs. Furthermore, we find that more compact clusters exhibit a stronger microlensing impact. Consequently, we conclude that clustering makes the existence of a significant population of BHs in the substellar to intermediate mass range even more unlikely.}

\maketitle

\section{Introduction} 
The nature of Dark Matter (DM) is one of the most challenging questions of modern Physics and still remains an open question almost a century after its discovery \citep[see, e.g.,][]{Swart:2017}. There are several well-motivated candidates in particle physics and Astrophysics to be the constituents of DM \citep[for an extensive review, see][]{2022JHEAp..34...49A}, but the lack of detection of weakly interacting elementary particles \citep{Chapline:2016}\footnote{Other particles like ultralight bosons (fuzzy dark matter) or sterile neutrinos are nowadays favored over WIMPs.} and the recent findings of the Laser Interferometer Gravitational-Wave Observatory \citep[LIGO, see][and references therein]{arXiv:2108.01045,arXiv:2010.14533}, have renewed the hypothesis that Dark Matter may consist of MAssive Compact Halo Objects (MACHOs). In particular, primordial black holes (PBHs) in the \SIrange{1}{1000}{\Msun} mass range are a favored candidate, as their abundance is not constrained by the baryon fraction determined in the analysis of the Cosmic Microwave Background by the Planck-collaboration \citep{PhysRevD.94.083504,2016A&A...594A..13P}. \citet{2009ApJ...706.1451M} found that quasar microlensing provides an excellent method to analyze the fraction of compact objects in lens galaxies \citep[an extension of the MACHO experiment,][to the extragalactic domain]{10.1086/309512} and used this method to check whether the existence of primordial black holes in this mass range is plausible \citep{2017ApJ...836L..18M}. They found that the observed microlensing is consistent with the normal stellar population. This conclusion has been supported by a later analysis considering a bimodal mass spectrum to take into account a mixed population of stars and black holes (BHs) \citep{2022ApJ...929..123E} and X-ray microlensing data to explore the sub-stellar mass range \citep{2023ApJ...954..172E}. 
On the other hand, \citet{Carr:2023} compiled a review of evidence in favor of a significant population of stellar-mass primordial black holes. For stellar-mass PBH, in particular \citet{2020A&A...643A..10H,2020A&A...633A.107H,  2022MNRAS.512.5706H} claim detection of a significant fraction of mass in the form of these compact objects in the halos of galaxies and/or clusters.

These previous studies are based on the assumption of a locally homogeneous spatial distribution of the individual microlenses for each considered population \citep{1999AAS...195.4802B,2015ApJ...806..251J}. However, theories about the origin and evolution of PBHs \citep{2017arXiv171004694G,2022JCAP...08..035G,PhysRevD.105.083520}  predict that primordial black holes are clustered, thereby inducing specific microlensing properties. In fact,  at a sufficient distance from a compact group of BHs, light rays can experience a deflection as if the cluster were a single object whose mass is equal to the sum of the individual masses. This would modify the spatial scale of microlensing variations, which is proportional to $\sqrt{M_\mathrm{lens}}$. Consequently, for finite-size sources, the strength of microlensing can have a trend to increase with the cluster mass and, more importantly, the minimum size that a source must have to average and wash out the spatial variations of microlensing will also increase. This invalidates the common practice of using the Broad Line Region (BLR) or the dust torus as a region unaffected by microlensing.

Thus, the objective of this work is to estimate the abundance of clustered black holes according to the observed microlensing in a sample of 12 lensed quasar systems. We will assume that each cluster contains \numrange{300}{3000} black holes and has collapsed to parsec scales \citep{2017arXiv171004694G,2022JCAP...08..035G,PhysRevD.105.083520}. Motivated by the masses of the black holes detected by LIGO, we fix the mass of individual black holes to $30\,\Msun$. We note that our choice of emission regions makes our analysis virtually immune to contamination by stellar-mass lenses, which includes both stars in the host galaxy and stellar-mass black holes of the type analyzed by \citet{Carr:2023}.

The article is organized as follows. In Section 2 we present the data and macro lens models collected from the literature. Section 3 is devoted to microlensing simulations and Bayesian analysis using lens models to define the zero microlensing baseline. In section 4 we compute the expected abundances of PBH clusters. In section 5 the limitations and consequences of our results are discussed. Finally, the main conclusions are summarised in section 6.
 
\begin{figure}[t]
    \centering
    \includegraphics[width=\linewidth]{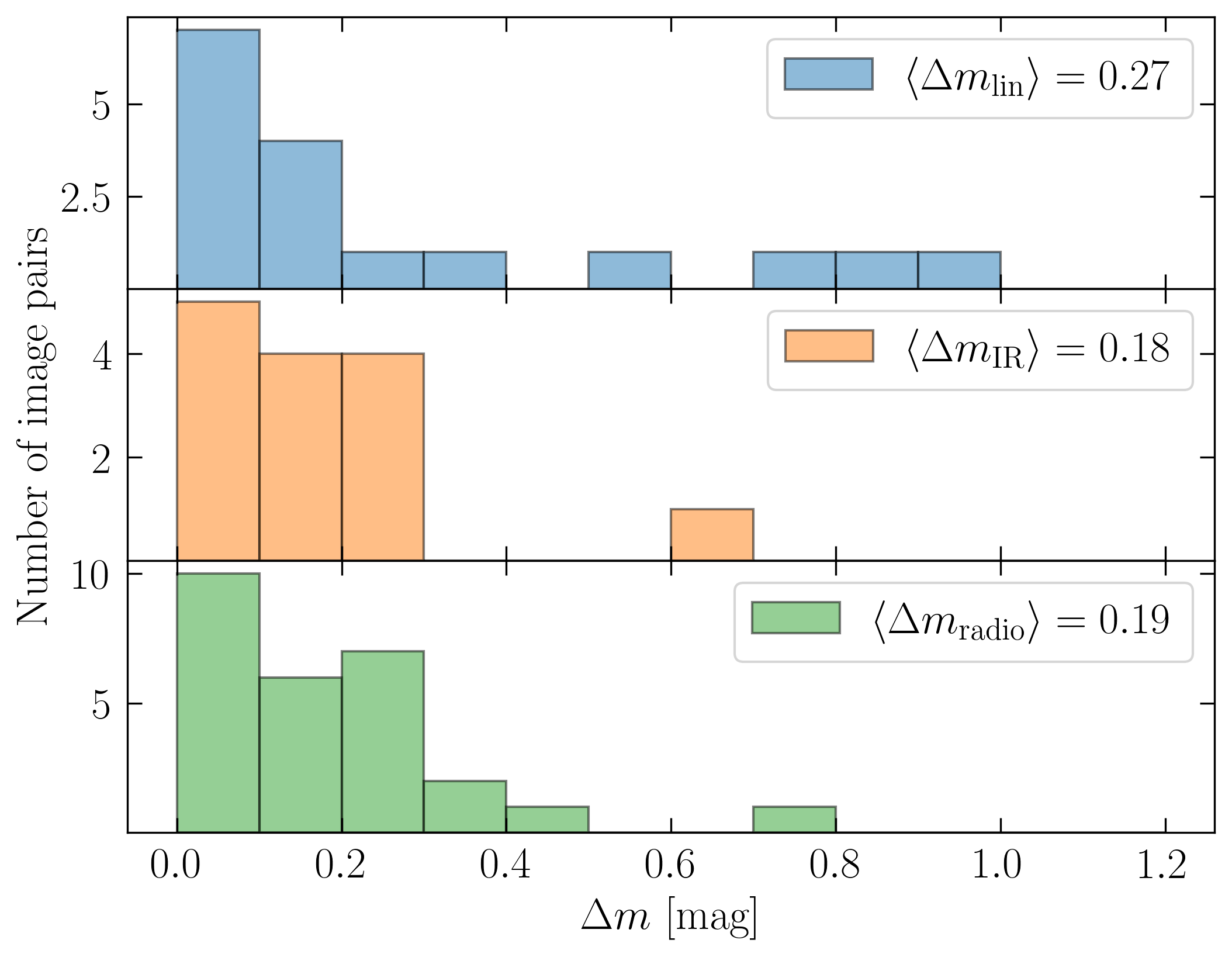}
    \caption{Histogram of the observed microlensing magnifications for the quasar systems. The x-axis represents the absolute value of observed magnification; the y-axis represents how often a certain magnification was observed in total. The three types of flux (line emission, infrared emission, and radio emission) are denoted by the colors green, orange, and blue, respectively. }
    \label{fig:observed_magnification_histograms}
\end{figure}

\section{Data}

Our study focuses on comparing observed flux ratios in lens systems with predictions formulated by models of their smooth mass distributions. Essential to our study are models that do not use
measured flux ratios as model constraints, yet they remain sufficiently informative to yield valuable forecasts. Consequently, our analysis is limited to lens systems exhibiting quadruply imaged source quasars, colloquially termed `quads'. 

As we aim to study the effect produced by massive clusters of PBHs, we would like our observations to be virtually free of the effect of microlensing by stars. Therefore our selection criterion encompasses those systems previously observed at wavelengths that are emitted by large source regions insensitive to stellar microlensing. This includes the spectral lines in visible light or near-infrared emanating from the quasar's broad-line region (BLR), mid-infrared light radiating from the dusty torus, or radio waves originating from the jet or other expansive regions related to the accretion phenomenon. Given the insensitivity of mid-infrared and radio wavelengths to extinction in the interstellar medium, these observations are presumed to provide accurate flux ratios between the multiple quasar images.

For our research, mass models from the studies of \citet{Sluse:2012,Schechter:2014}
were preferentially adopted where applicable. The publication by \citet{Schechter:2014} is particularly salient as it already provides values for the convergence ($\kappa$) and shear ($\gamma$) of the lens potential at the precise locations of the quasar images, simplifying the production of magnification maps. In the cases of B 0712+472 and B 1555+375, models including an exponential disk for the lens were used following the findings of \citet{Hsueh:2016} and \cite{Hsueh:2017}. These simple models are prioritized irrespective of the availability of other, even newer, more sophisticated models from diverse sources\footnote{The more advanced models were mostly constructed to measure the Hubble constant $H_0$ to percent-level precision \citep[compare][and references therein]{2021MNRAS.503.1096D}. As neither our flux ratio measurements nor our microlensing models have that precision, we believe including these much more complex models is not necessary.}. Our selection of lens systems, shown in Table \ref{tab:lenses_properties}, includes details such as predicted and observed flux ratios in various wavelengths as well as flux-ratio anomalies represented in magnitudes. Notes on individual systems can be found in Appendix \ref{app:indiv-systems}.

\begin{table*}[ht]

  \centering
  \caption{Observed flux ratios for our lens samples}
  \begin{threeparttable}
\begin{tabular}{cccccccccc}
Object & Model (constraints) & Ratios & Model & Line & IR & Radio & $\Delta m_{l i n}$ & $\Delta m_{I R}$ & $\Delta m_{\text {radio }}$ \\
\multirow[t]{3}{*}{ B 0128+437\tnote{1,3}} & $\mathrm{SIE}+\gamma$ & $\mathrm{B} / \mathrm{A}$ & 0.72 & - & - & 0.58 & - & - & 0.23 \\
& $\left(\vec{x}_{i}, \vec{x}_{G 1}\right)$ & $\mathrm{C} / \mathrm{A}$ & 0.40 & - & - & 0.52 & - & - & -0.28 \\
& & $\mathrm{D} / \mathrm{A}$ & 0.48 & - & - & 0.51 & - & - & -0.07 \\
\multirow[t]{3}{*}{MG 0414+0534\tnote{2,4}} & $\mathrm{SIE}+\gamma+\mathrm{SIS}$ & $\mathrm{A} 2 / \mathrm{A} 1$ & 1.03 & - & 0.90 & 0.90 & - & 0.14 & 0.14 \\
& $\left(\vec{x}_{i}, \vec{x}_{G 1}, \vec{x}_{G X}\right)$ & B/A1 & 0.29 & - & 0.36 & 0.37 & - & -0.25 & -0.28 \\
& & $\mathrm{C} / \mathrm{A} 1$ & 0.15 & - & 0.12 & 0.15 & - & 0.22 & 0.00 \\
\multirow[t]{3}{*}{HE 0435-1223\tnote{2,5,6}} & $\mathrm{SIE}+\gamma$ & $\mathrm{A} / \mathrm{C}$ & 0.94 & 0.97 & 1.71 & 1.05 & -0.03 & -0.64 & -0.12 \\
& $\left(\vec{x}_{i}, \vec{x}_{G 1}\right)$ & $\mathrm{B} / \mathrm{C}$ & 1.02 & 0.98 & 0.99 & 0.77 & 0.04 & 0.03 & 0.31 \\
& & $\mathrm{D} / \mathrm{C}$ & 0.61 & 0.66 & 0.81 & 0.47 & -0.08 & -0.30 & 0.28 \\
\multirow[t]{3}{*}{RX J0911+0551\tnote{2,6}} & $\mathrm{SIE}+\gamma+\mathrm{SIS}$ & $\mathrm{B} / \mathrm{A}$ & 1.85 & - & - & 1.98 & - & - & 0.08 \\
& $\left(\vec{x}_{i}, \vec{x}_{G 1}\right)$ & $\mathrm{C} / \mathrm{A}$ & 0.85 & - & - & 0.73 & - & - & -0.17 \\
& & $\mathrm{D} / \mathrm{A}$ & 0.34 & - & - & 0.35 & - & - & 0.03 \\
\multirow[t]{3}{*}{ PG 1115+080\tnote{2,8}} & $\mathrm{SIS}+\mathrm{SIS}$ & $\mathrm{A} 2 / \mathrm{A} 1$ & 0.95 & 1.00 & 0.93 & - & -0.05 & 0.03 & - \\
& $\left(\vec{x}_{i}, \vec{x}_{G 1}\right)$ & $\mathrm{B} / \mathrm{A} 1$ & 0.169 & - & 0.16 & - & - & 0.06 & - \\
& & $\mathrm{C} / \mathrm{A} 1$ & 0.256 & - & 0.21 & - & - & 0.22 & - \\
\multirow[t]{2}{*}{ RXS J1131-1231\tnote{2,9}} & $\mathrm{SIE}+\gamma$ & $\mathrm{A} / \mathrm{B}$ & 1.62 & 1.63 & - & - & -0.01 & - & - \\
& $\left(\vec{x}_{i}, \vec{x}_{G 1}\right)$ & $\mathrm{C} / \mathrm{B}$ & 0.94 & 1.19 & - & - & -0.26 & - & - \\
\multirow[t]{3}{*}{ B J1422+231\tnote{2,3,10}} & $\mathrm{SIE}+\gamma$ & $\mathrm{B} / \mathrm{A}$ & 1.18 & 1.11 & 1.06 & 1.06 & 0.07 & 0.11 & 0.11 \\
& $\left(\vec{x}_{i}, \vec{x}_{G 1}\right)$ & $\mathrm{C} / \mathrm{A}$ & 0.62 & 0.54 & 0.61 & 0.55 & 0.15 & 0.02 & 0.13 \\
& & $\mathrm{D} / \mathrm{A}$ & 0.05 & 0.03 & - & 0.02 & 0.54 & - & 0.49 \\

\multirow[t]{3}{*}{ B 1608+656\tnote{1,11}} & $\mathrm{SIE}+\gamma+\mathrm{SIS}$ & $\mathrm{A} / \mathrm{B}$ & 1.95 & - & - & 2.04 & - & - & -0.05 \\
& $\left(\vec{x}_{i}, \vec{x}_{G 1}, \vec{x}_{G 2}\right)$ & $\mathrm{C} / \mathrm{B}$ & 0.81 & - & - & 1.04 & - & - & -0.27 \\
& & $\mathrm{D} / \mathrm{B}$ & 0.17 & - & - & 0.35 & - & - & -0.79 \\
\multirow[t]{3}{*}{ WFI J2033-4723\tnote{2,12}} & $\mathrm{SIE}+\gamma+\mathrm{SIS}$ & $\mathrm{A} 1 / \mathrm{B}$ & 1.56 & 1.48 & - & - & 0.06 & - & - \\
& $\left(\vec{x}_{i}, \vec{x}_{G 2}\right)$ & $\mathrm{A} 2 / \mathrm{B}$ & 0.92 & 1.05 & - & - & -0.14 & - & - \\
& & $\mathrm{C} / \mathrm{B}$ & 0.62 & 1.39 & - & - & -0.88 & - & - \\
\multirow[t]{3}{*}{ Q 2237+0305\tnote{13,14}} & $\mathrm{SIE}+\gamma$ & $\mathrm{B} / \mathrm{A}$ & 0.89 & 0.81 & 0.97 & 1.08 & 0.10 & -0.09 & -0.21 \\
& $\left(\vec{x}_{i}, \vec{x}_{G 1}\right)$ & $\mathrm{C} / \mathrm{A}$ & 0.45 & 0.88 & 0.51 & 0.55 & -0.73 & -0.13 & -0.22 \\
& & $\mathrm{D} / \mathrm{A}$ & 0.82 & 1.09 & 0.92 & 0.77 & -0.31 & -0.13 & 0.07 \\
\multirow[t]{3}{*}{ B 1555+375\tnote{3,15}} & $\mathrm{SIE}+\mathrm{Exp.\,disc}$ & $\mathrm{B} / \mathrm{A}$ & 0.61 & - & - & 0.62 & - & - & -0.02 \\
& $\left(\vec{x}_{i}, \vec{x}_{G 1}\right)$ & $\mathrm{C} / \mathrm{A}$ & 0.45 & - & - & 0.51 & - & - & -0.14 \\
& & $\mathrm{D} / \mathrm{A}$ & 0.12 & - & - & 0.09 & - & - & 0.31 \\
\multirow[t]{3}{*}{ B 0712+472\tnote{3,16}} & $\mathrm{SIE}+\gamma+\mathrm{Exp.\,disc}$ & $\mathrm{B} / \mathrm{A}$ & 0.82 & - & - & 0.84 & - & - & -0.03 \\
& $\left(\vec{x}_{i}, \vec{x}_{G 1}\right)$ & $\mathrm{C} / \mathrm{A}$ & 0.43 & - & - & 0.42 & - & - & 0.03 \\
& & $\mathrm{D} / \mathrm{A}$ & 0.086 & - & - & 0.082 & - & - & 0.05 \\
\hline
\end{tabular}    
\begin{tablenotes}
\item[1] Model ratios from \citet{Sluse:2012}

\item[2] Model ratios from \citet{Schechter:2014}.


\item[3] Radio flux ratios from \citet{2003ApJ...595..712K}. 

\item[4] Mid-IR flux ratios from \citet{Minezaki:2009}, radio flux ratios from \citet{Rumbaugh:2015}.

\item[5] emission line flux ratios from \citet{Nierenberg:2017}, Flux ratios from \citet{Fadely:2012}

\item[6] Radio flux ratios from \citet{Jackson:2015}.

\item[7] Ly $\alpha$ flux ratios from \citet{Keeton:2006}.

\item[8] Optical emission line flux ratios from \citet{Popovic:2005}. Mid-IR data from \citet{Chiba:2005}.

\item[9] [OIII] emission line flux ratios from \citet{Sugai:2007}.

\item[10] Optical emission line flux ratios from \citet{Impey:1996}. Mid-IR data from \citet{Chiba:2005}.

\item[11] Radio flux ratios from \citet{Fassnacht:1999}.

\item[12] Optical emission line flux ratios from \citet{Morgan:2004}.

\item[13] [OIII] emission line flux ratios from \citet{Wayth:2005}. Mid-IR and model flux ratios from \citet{Vives-Arias:2016}. Radio flux ratios from \citet{Falco:1996}.

\item[14] Model ratios from \cite{Vives-Arias:2016}

\item[15] Model ratios from \cite{Hsueh:2016} 

\item[16] Model ratios from J. Cohen (priv. comm.)

\end{tablenotes}
  \end{threeparttable}
  \label{tab:lenses_properties}
\end{table*}

The observed flux-ratio anomalies for all selected lens systems are shown in Fig.~\ref{fig:observed_magnification_histograms}. No significant differences are visible between the three types of flux considered (line, infrared, and radio emissions), even if they come from regions of different sizes. 

\section{Methods}
\subsection{Generating clustered black holes}

\begin{figure}[t!]
    \centering
    \includegraphics[width=\linewidth]{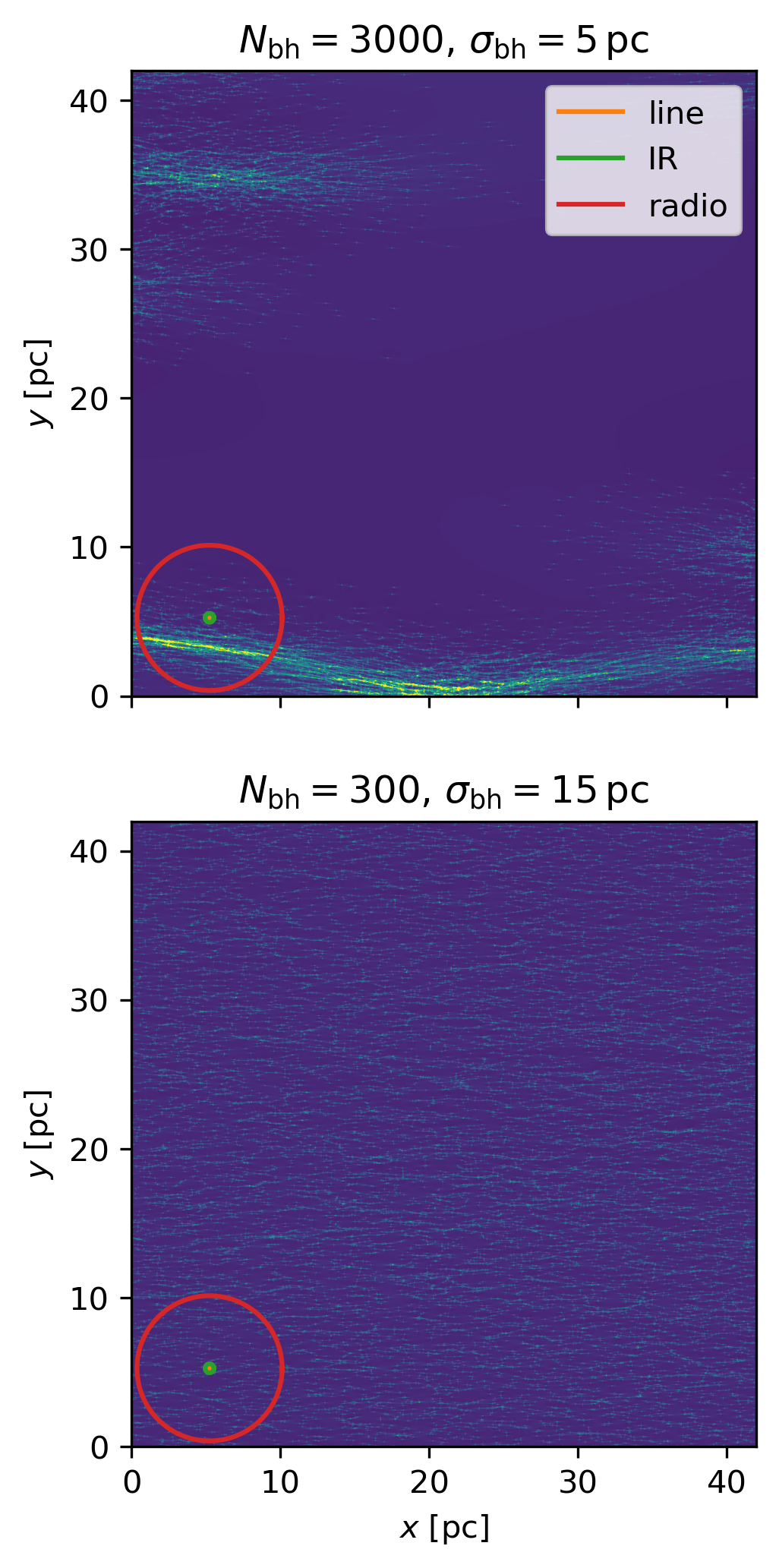}
    \caption{Two example magnification maps for the image C of {B~1555}. In both cases, the fraction of microlenses is $\alpha=0.2$. The source sizes for line, infrared, and radio emission are indicated as orange, green, and red circles respectively. The top panel shows that when many black holes are contained in a very dense cluster, the magnification map is dominated by the features caused by the clusters themselves (notice the flocculent giant caustic). The magnification distribution in the lower panel, on the contrary, appears closer to that of a uniform lens population.}
    \label{fig:magnification_maps_clustered_bh}
\end{figure}
\begin{figure}[th!]
    \centering
    \includegraphics[width=\linewidth]{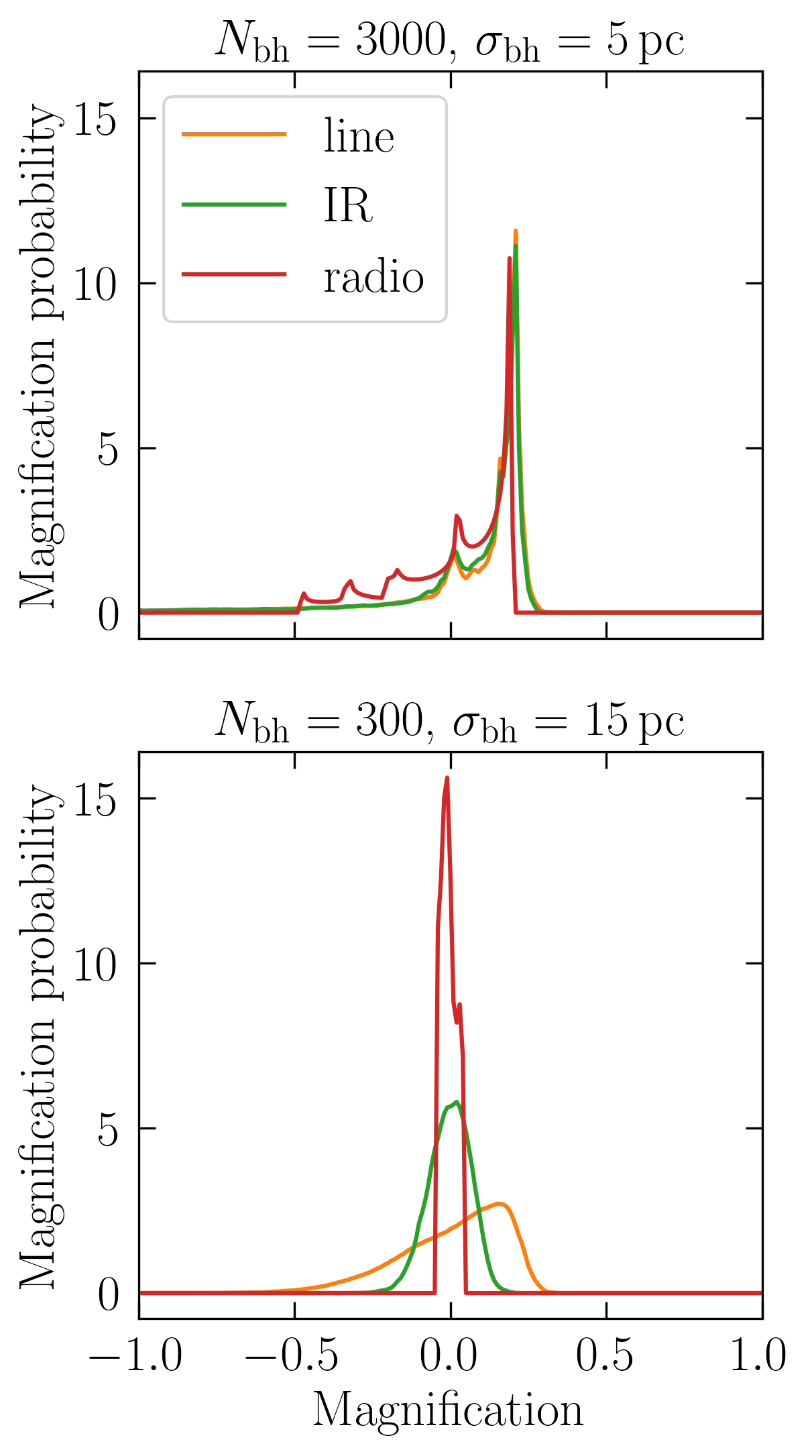}
    \caption{Histograms of the two example magnification maps for the image C of {B~1555}. In both cases, the fraction of microlenses is $\alpha=0.2$. One can see that for the upper case (more massive clusters of less spatial extent) the magnification pattern is more structured and behaves very similarly for the three different source sizes because the relative impact of the differences in size is significantly reduced in the presence of the much larger features induced by the clusters acting like giant pseudo-particles.}
    \label{fig:magnification_maps_clustered_bh_histograms}
\end{figure}

    We have no prior knowledge of the radial profile of black hole clusters. After performing several tests based on magnification map computations using different radial profiles (Gaussian vs top-hat), we found that the microlensing magnification statistics are rather insensitive to the radial profile, so we adopted a Gaussian profile for the rest of the present work. For the cluster sizes, we follow \cite{2022JCAP...08..035G} and \cite{PhysRevD.105.083520} and we have taken $\sigma_{\rm bh}$ equal to \SIlist{5;10;15}{pc}. We also take as variables in the simulations the number of black holes per cluster, $N_{\rm bh}$,  for which we explored the range \numlist{300;1000;3000} \citep[cf.][]{2022JCAP...08..035G}, and the fraction of mass in microlenses (i.e., in the BH clusters), $\alpha=\kappa_{\rm bh}/\kappa$, which can take values in \numlist{0.0125;0.025;0.05;0.1;0.2;0.3;0.4;0.6}. Larger values of $\alpha$ were explored in the initial stage but were excluded from the final study due to the low likelihood of these models. The masses of the black holes are fixed to \SI{30}{\Msun}, which is in rough agreement both with the LIGO-observations and with \citet{2017arXiv171004694G}. In reality, the mass function is expected to be more complex \citep[][suggested a lognormal profile]{2017arXiv171004694G}, but single epoch microlensing is only weakly sensitive to the shape of the mass spectrum \citep{2007MNRAS.376..263C}; the only important quantity being the geometric mean of the mass distribution \citep{2020ApJ...904..176E}. Our model parameter space therefore consists of $3\times3\times8=72$ models.

    As we use flux ratio measurements from broad lines, infrared, and radio, whose emission regions have different sizes, we must model these different sources separately. For the size of the emission regions, we adopted a Gaussian luminosity profile \citep[the statistics of microlensing magnification of a source is not sensitive to the source brightness profile, but only to its half-light radius, cf.][]{2005ApJ...628..594M}. We choose half-light radii corresponding to $\sigma_\mathrm{line} = \SI{100}{lt-day}$  \citep[cf.][]{2019ApJ...887...38G}, $\sigma_\mathrm{IR}=\SI{1}{ly}$ \citep[cf.][]{2023arXiv230212437L} and $\sigma_\mathrm{radio}=\SI{5}{pc}$ for the lines, IR and radio respectively. While the sizes for the BLR and IR are reasonably well established (see references above), the size of the radio emission in radio-quiet quasars (RQQ) is much more uncertain, with a fraction of the emission possibly coming from a very extended region related to star formation. Nevertheless, \cite{2017MNRAS.468..217W} have shown that the radio emission in a vast majority of quasars is accretion dominated, and Very Long Baseline Interferometry (VLBI) images of RQQs by \cite{2023MNRAS.518...39W} show that 5pc can be used as a bona fide rough estimate for the compact radio emission in quasars. The similar width of the three microlensing magnification histograms in Fig.~\ref{fig:observed_magnification_histograms} supports either a not-too-dissimilar size for the three emitting regions considered or at least indirect evidence that the sizes of the three regions are quite small as compared with the Einstein radius/scale of the microlensing magnification patterns\footnote{A third, possible but more unlikely explanation would be that the measured flux ratios are an artifact of either very uncertain measurements or incorrect lens models.}.

    From the descriptions of the individual systems (see Appendix \ref{appena}), we note that the sizes of some radio sources appear to be significantly larger than the 5 pc assumed in this study. In some cases, these quoted source sizes do not concern the half-light radius of the source (which is the essential quantity for microlensing), so while part of their emission might originate from an extended jet, it does not invalidate our assumption. Adapting a different source size for each lens system would indeed be the ideal solution, but we lack the data to perform such a study (and simulating sources of $\sim$300 pc is computationally prohibitive). Moreover, if the effective radio sizes were systematically larger than our assumption, then we would see a significantly different microlensing behavior (purportedly smaller) from the one of the IR  or BLR emission, which is not supported by the observations (see Fig.~\ref{fig:observed_magnification_histograms}). 
    
    For each point in our parameter space ($\sigma_{\rm bh}$, $N_{\rm bh}$, $\alpha$), we calculate 25 magnification maps (see below), which are then convolved with the corresponding source luminosity profile for each of the three different emission sizes. Finally, we obtain the histogram for each of the convolved map sets (previously normalized to the expected mean magnification).

\subsection{Magnification maps}

Due to the large source sizes considered in the present work ($\sim 0.1-5$ pc) and also to the large extent of the PBH clusters (up to $\sim 15$ pc), the needed magnification maps must cover a very large region of both source and lens planes. We calculate magnification maps containing $2000 \times 2000$ pixels$^2$ and fix a pixel size in the magnification map to be $25\times 25\,\mathrm{lightdays}^2$. This leads to magnification maps of approx. $\SI{42}\times\SI{42} {pc}^2$. To generate the magnification maps, an even larger region in the lens plane must be populated with lenses, which can consequently amount to very large numbers (up to a few times $10^8$ lenses in the most extreme cases). To be able to calculate these computationally heavy magnification maps in a reasonable amount of time, we have taken benefit from the new FMM-IPM algorithm described in \cite{2022ApJ...941...80J} which combines the inverse polygon mapping algorithm \citep{2011ApJ...741...42M,2006ApJ...653..942M} designed to efficiently construct low noise magnification maps, with the fast multipole method \citep{1987JCoPh..73..325G} for ray deflection calculation, which is very efficient when a large number of lenses is involved, as is frequently the case in the present work.
For each of the 48 quasar images and each choice of the model parameters $\Nbh$, $\Rbh$, and $\alpha$, we generate 25 magnification maps to reduce statistical noise.
The present work therefore needed $3\times3\times8\times48\times25=86400$ magnification maps with a typical number of $\sim 10^6$ lenses, which constitutes a huge computational workload. In the case of clustered black holes, averaging over many magnification maps is crucial: The shot noise is significantly increased, as the dominant lensing contribution stems from the clusters of BH, not the individual lenses themselves. We found that the average magnification of a single magnification map scatters by around 10\% around the expected one (for the highest values of $\alpha$), with some extreme cases reaching up to 30\% deviation. The average magnification of all magnification maps, however, is within 2\% of the theoretical one.
We show two examples of magnification maps in Fig.~\ref{fig:magnification_maps_clustered_bh} and their respective magnification histograms in Fig.~\ref{fig:magnification_maps_clustered_bh_histograms}.

\subsection{Baseline determination}

    To compare the simulations with the observations we need to separate the magnification by microlensing from the PBH clusters from the magnification by the macro lens. This is done by establishing a zero microlensing baseline of magnification. Usually, when addressing microlensing by stellar-like objects, the emission coming from a large region that is mostly unaffected by microlensing (e.g. the broad-line emission region or the emission from the dust torus in the mid-IR) is used for that purpose \citep[cf.][]{2009ApJ...706.1451M, Vives-Arias:2016}.

    When many lenses are concentrated in clusters of small physical extent (as in the $\Nbh=3000$, $\sigma_\mathrm{BH}=5\,\mathrm{pc}$ case of Fig.~\ref{fig:magnification_maps_clustered_bh} and \ref{fig:magnification_maps_clustered_bh_histograms}), we find that the magnification maps of clustered lenses are dominated by the huge caustics caused by the clusters of black holes: 
    to a light ray passing outside of a cluster, the whole cluster acts as a single lens with the combined mass of all the black holes. As the fraction of area taken by the clusters is relatively small, the features caused by the clusters dominate over the ones caused by single black holes. This strongly aggravates the determination of a baseline. In the standard scenario, with microlenses of typical stellar masses,
    the broad line emission region is much larger than the Einstein radius of the microlenses, so that any microlensing effect is {effectively} `washed out', i.e. the BLR is {virtually} insensitive to microlensing. However, in the present scenario, we are dealing with black holes whose masses are about \num{100} times larger than the ones of average stars. A cluster of \num{1000} of those black holes now induces features that are about \num{300} times larger and therefore, can still significantly affect the BLR or the dust torus.

    Considering this, we can state that microlensing experiments that compute the baseline using the broad-line emission region (or even the radio emission) cannot detect any effects of clustered primordial black holes, as the baseline can be also significantly (de)magnified by microlensing.  
    Therefore, previous results that exclude the existence of primordial black holes due to lack of observed microlensing using such baselines \citep[like][]{2017ApJ...836L..18M} do not apply to the case of clustered primordial black holes. Consequently, a new study using a different method of baseline determination is necessary to exclude these. In this work, we therefore use the magnification predicted by the lens model as the baseline. While this method avoids contamination of the baseline by micro/milli-lensing, potential inaccuracies in the lens model give rise to uncertainties in the observed magnification, which we will discuss in the next section.

\subsection{A Bayesian analysis of the microlensing signal}
\label{sec:bayesian_analysis}
\begin{figure}
    \centering
    \includegraphics[width=\linewidth]{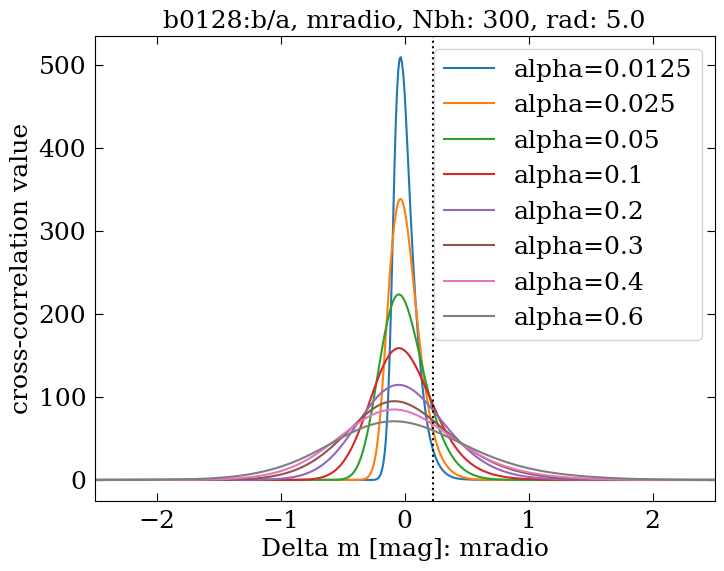}
    \caption{Example cross-correlation of magnification histograms. The colored lines describe the probability of observing a magnification difference of $\Delta m$ between the images a and b of the B 0128+437 lens system for the model with $N_\mathrm{bh}=300$ and $\sigma_\mathrm{bh}=5$ pc. The dashed line represents the actually observed difference. We note that the smoothing of $\Delta\, m=0.24$ to account for observational/model errors has already been applied to these histograms.}
    \label{fig:cc_mag_histograms}
\end{figure}
     To compare our models to observations, we calculate the expected magnification for each image according to
     \begin{equation}
         \mu = \frac{1}{(1-\kappa-\gamma)\,(1-\kappa+\gamma)},
     \end{equation}
     where $\kappa$ and $\gamma$ denote the local convergence and shear, according to the models referenced in Tab.~\ref{tab:lenses_properties}. We then divide the measured flux by this baseline and compute the difference (in magnitudes, {$\Delta m$}) between the brightness of two separate images {$j$ and $k$}. In our simulations, this corresponds to a cross-correlation of the magnification histograms{, $P(m)$,} of the two images {$P(\Delta m)=\int P_j(m)\,P_k(m+\Delta m)\, \mathrm{d}m$} \citep[cf.][]{2009ApJ...706.1451M}. {To take into account the uncertainties in the measured value $\sigma_{\Delta m}$, this cross-correlated histogram must be convolved with a Gaussian kernel of width $\sigma_{\Delta m}$. From this probability distribution for $\Delta m$, we can calculate the likelihood of observing a given $\Delta m_i$ for image pair $i$, for a certain model parametrized by $\alpha, N_{bh}, \sigma_{bh}$: $P(\Delta m_i|\alpha, N_{bh}, \sigma_{bh})$. One such {likelihood} can be seen in Fig.~\ref{fig:cc_mag_histograms}.} 
     
     To estimate these errors in the difference in magnitudes, $\sigma_{\Delta m}$, it is essential to consider both the uncertainty in the flux-ratio measurement (which depends on two single flux measurements) and the uncertainty in the flux ratio predicted by the lens model. To make a conservative estimate of the uncertainty in the experimental flux ratios, we can determine the standard deviation of the magnitude differences corresponding to the same image pair in different bands (emission lines, infrared, and radio). We have
     calculated standard deviations for the 12 image pairs in Table \ref{tab:lenses_properties} that have measurements of the flux ratios in more than one band. The mean of the standard deviations is 0.14 magnitudes\footnote{It is worth noting that this value is close to the 0.11 magnitudes of the average difference between mid-IR and emission line flux ratios found by \citep{2009ApJ...706.1451M}.}, corresponding to a typical error of approximately 14\% in the flux-ratio measurements, providing a positive cross-check of data reliability.
     The uncertainty in the models is likely significantly larger and, unfortunately, much more challenging to evaluate. We will adopt an uncertainty of $\sim$20\%  \citep[likely a lower limit;][use this error level to compare flux-ratios from models and observations, finding that the differences exceed the statistical expectations]{2019MNRAS.483.5649S}. By combining both error sources, we obtain an overall estimate of the error $\sigma_{\Delta m} = 0.24$ mag. This is likely to be still an underestimation of the true error since we did not consider potential secondary effects like millilensing by subhaloes. Underestimating this uncertainty may lead to an artificial overestimate of the magnification anomalies (by considering as a true signal what is only compatible with a weaker but noisy signal) and, consequently, to an overestimate of the likelihood of the existence of a population of clustered PBHs. In Sect.~\ref{sec:results:small_errors}, we nevertheless analyze this effect and show the results when the uncertainties in the model are ignored by assuming $\sigma_{\Delta m} \sim 0.14$ mag. We note that, both by theoretical reasoning and numerical results in the following sections, a clear pattern is established in which smaller uncertainties lead to larger possible abundances of PBH microlenses. This means that other choices of error estimation that lead to a total uncertainty larger than 0.14 mag (for example disregarding flux measurement uncertainties, but keeping model uncertainties) will not allow higher PHB abundances than the ones discussed in Sect.~\ref{sec:results:small_errors}.
    
Finally, we extract a {posterior probability distribution} for the chosen parameters {given the observation $\Delta m_i$, $P(\alpha, N_{bh}, \sigma_{bh}| \Delta m_i)$} by applying Bayes' Theorem to the resulting 
{likelihood as $P(\alpha, N_{bh}, \sigma_{bh}|\Delta m_i)=P(\Delta m_i|\alpha, N_{bh}, \sigma_{bh}) P(\alpha, N_{bh}, \sigma_{bh})$)}. 
{We will assume here non-informative priors for all model parameters $\alpha, N_{bh}, \sigma_{bh}$.}

    Unfortunately, the large computational requirement of creating \num{1175} magnification maps for one point in parameter space, combined with the $\gtrsim 10^6$ required microlenses per magnification map, prevents us from performing an analysis using a Markov Chain Monte-Carlo (MCMC) framework. However, since our parameter space is only three-dimensional, we believe that a coarse sampling of these three dimensions is sufficient to explore the parameter space at this stage.

    For this analysis, we restrict ourselves to a uniform prior of $\alpha$, allowing values between 0.0125 and 0.6.

\section{Results}
\begin{figure*}
        \centering
        \includegraphics[width=\linewidth]{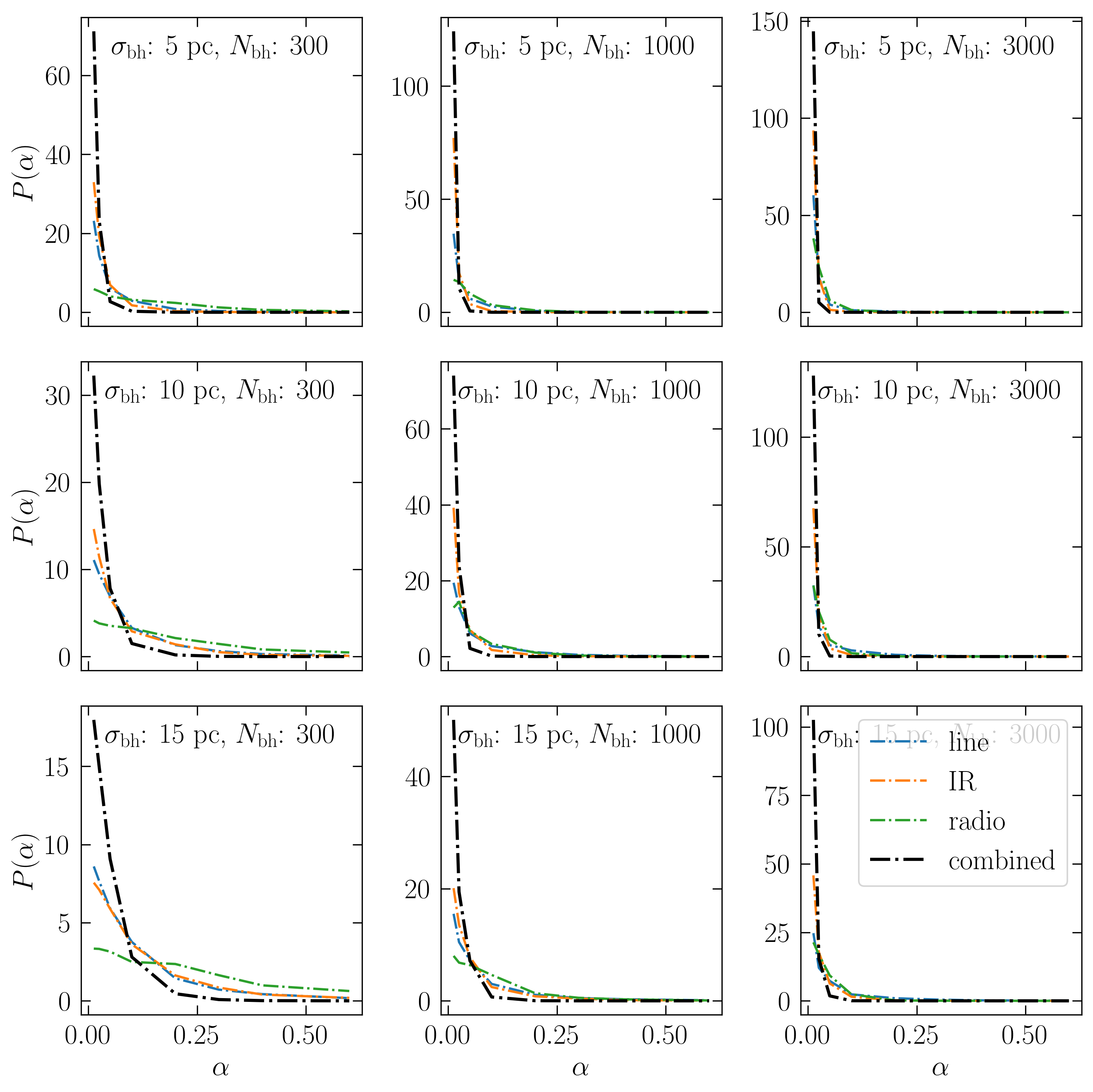}
        \caption{Probability distributions for the fraction of mass in microlenses $\alpha$. For each panel, the choice of $\Nbh$ and $\Rbh$ is fixed; the x-axis parametrizes the value of $\alpha$. The colored lines correspond to the probability distribution if only one type of flux (lines, infrared, or radio emissions) is considered. The black line corresponds to the combined probability distribution using all the available fluxes.}
        \label{fig:alpha_microlensing}
    \end{figure*}

    In this section, we present the results of our Bayesian analysis outlined in Sect.~\ref{sec:bayesian_analysis}. We note that our analysis is sensitive to our choice of modeling and measurement errors, where smaller errors tend to favor higher dark matter fractions $\alpha$. We will therefore perform two analyses, one fiducial analysis (Sect.~\ref{sec:results:fiducial}) with our best estimate for modeling and measurement errors, and one analysis where we assume (unrealistically) small errors to ensure that our conclusions are robust even if we were to over-estimate our fiducial errors. We note that all probability distributions were normalized such that their integral over the prior range equals one. Therefore, the y-axis range differs substantially between plots. We advise the reader to be mindful of this when interpreting subsequent figures.
    
    \subsection{Fiducial analysis}
        \label{sec:results:fiducial}
        The combined posterior probability density function (PDF) {for the whole sample} is obtained from the conflation of the individual PDFs for each pair: {$P(\alpha, N_{bh}, \sigma_{bh}))=\prod_i P(\alpha, N_{bh}, \sigma_{bh}|\Delta m_i)$, where the product is evaluated over all observed image pairs}, and it is shown in Fig.~\ref{fig:alpha_microlensing}, together with the rest of the posterior PDFs for all the explored values of $N_{bh}$ and $\sigma_{bh}$. The first conclusion that can be drawn from this Figure is that the combined PDFs peak at zero for all the (explored) values of the parameters. The upper limits for the abundance of PBHs ($\alpha$) range from 0.07 (0.18) at 68\% (95\%) confidence level at the most favorable case ($N_{bh}=300$, $\sigma_{bh}=15\rm\, pc$) to 0.02 (0.03) (at $N_{bh}=3000$, $\sigma_{bh}=5\rm\, pc$). 
        Fig.~\ref{fig:alpha_microlensing} shows that the maximum allowed fraction of mass in PBH clusters decreases with increasing $N_{bh}$ and decreasing $\sigma_{bh}$. Interpreting clusters as pseudo-particles, this trend implies that for a given total mass, the likelihood of high microlensing magnification increases with the concentration of the mass in clusters. Consequently, the relatively moderate observed anomalies make the permitted fraction of mass in PBH clusters increase with the size of the clusters. Nevertheless, even in the limiting case of a uniform distribution of isolated BHs, we know from previous microlensing studies \citep{2022ApJ...929..123E} that the estimated abundance of PBHs in this mass range is negligible. The main conclusion of our study is then that clustering increases the probability for strong microlensing magnifications, and makes the existence of a significant fraction of Dark Matter in clustered PBHs less consistent with observations than a uniform distribution of lenses.
    
\begin{figure}
            \centering
            \includegraphics[width=\linewidth]{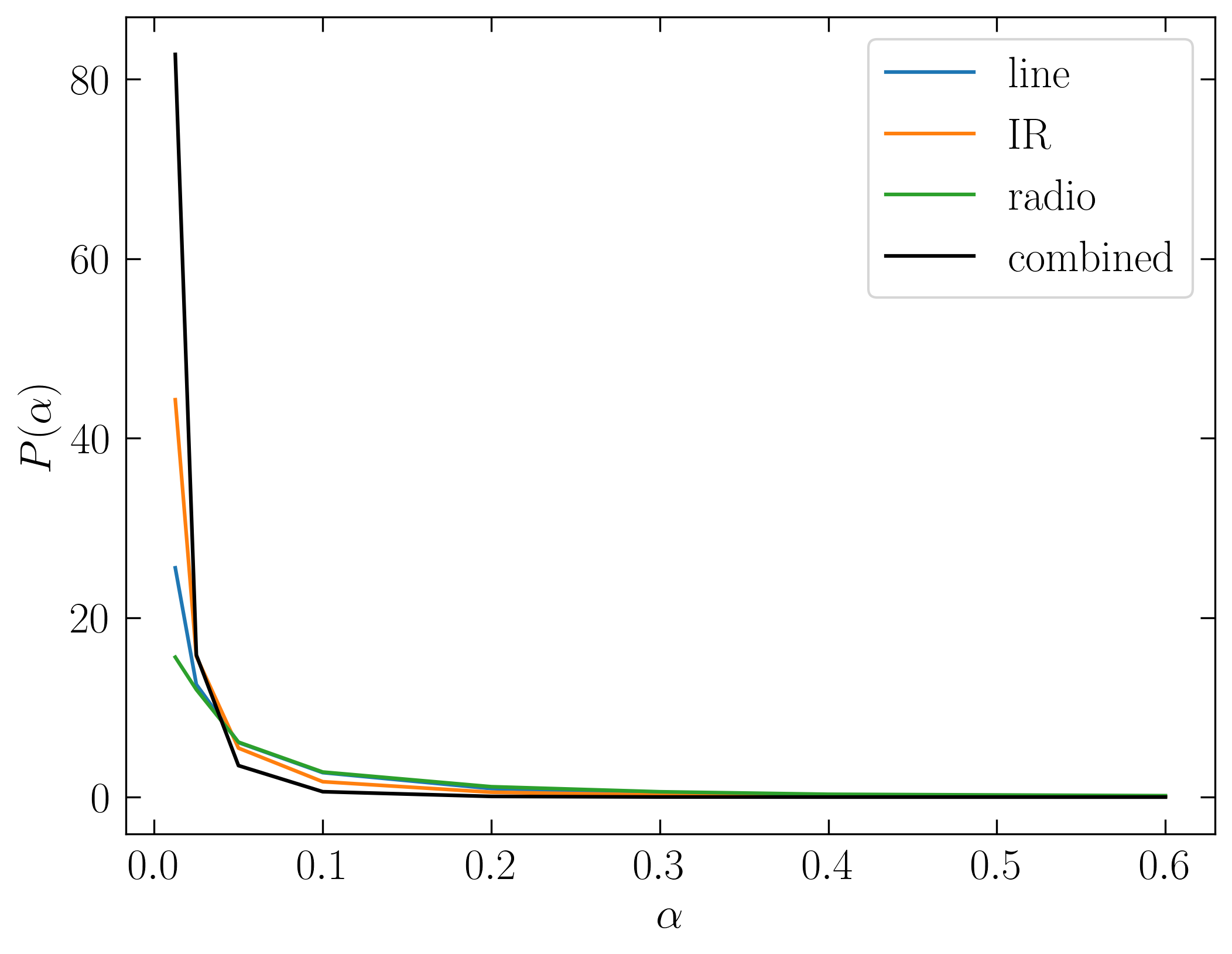}
            \caption{Probability distribution of fraction of mass in microlensing $\alpha$, here marginalized over all choices of $\Nbh$ and $\sigma_\mathrm{BH}$. }
            \label{fig:frac_alpha_marginalized}
        \end{figure}

        We can marginalize over our lack of knowledge in the cluster model parameters $N_{bh}$ and $\sigma_{bh}$ by taking the average of all magnification histograms\footnote{A better way to do this would certainly be to perform a proper marginalization via a Monte Carlo chain. However, our purpose in this work is to explore whether or not clustered PHBs are consistent with observations, not to derive precise constraints on them. As we have no way to set a physically motivated prior on $N_\mathrm{bh}$ or $\sigma_\mathrm{bh}$, and the computational load of an MCMC would be (close to) prohibitive, we believe this approach is satisfactory at this stage.} and produce a single PDF for the abundance of PBHs in clusters. This is shown in Fig.~\ref{fig:frac_alpha_marginalized}. From this PDF we can calculate an overall upper limit to the abundance of PBHs at 68\% (95\%) to be 0.03 (0.09). Fig.~\ref{fig:frac_alpha_marginalized} also shows the results for the broad lines, IR, and radio separately. The three sub-samples show very similar results, indicating that none of them introduces a significant bias in the final result.
    
        We are well aware that the results of our analysis depend on our model choices. While we are confident that all our analysis choices are justified, we can not fully exclude the hypothesis that a different set of model choices would lead to different posterior distributions and might ultimately allow a larger contribution of PBHs to dark matter. However, it is reasonable to suppose that, on average, any improvement of models will lead to a better prediction of the flux ratios with a subsequent reduction of the anomalies. The analysis choices most likely to make a difference are the values for the modeling and measurement uncertainties of the flux ratios between quasar images. We investigate this impact in Sect.~\ref{sec:results:small_errors}.

    \subsection{Quantifying the impact of model errors}

\begin{figure*}[ht]
        \centering
        \includegraphics[width=0.8\linewidth]{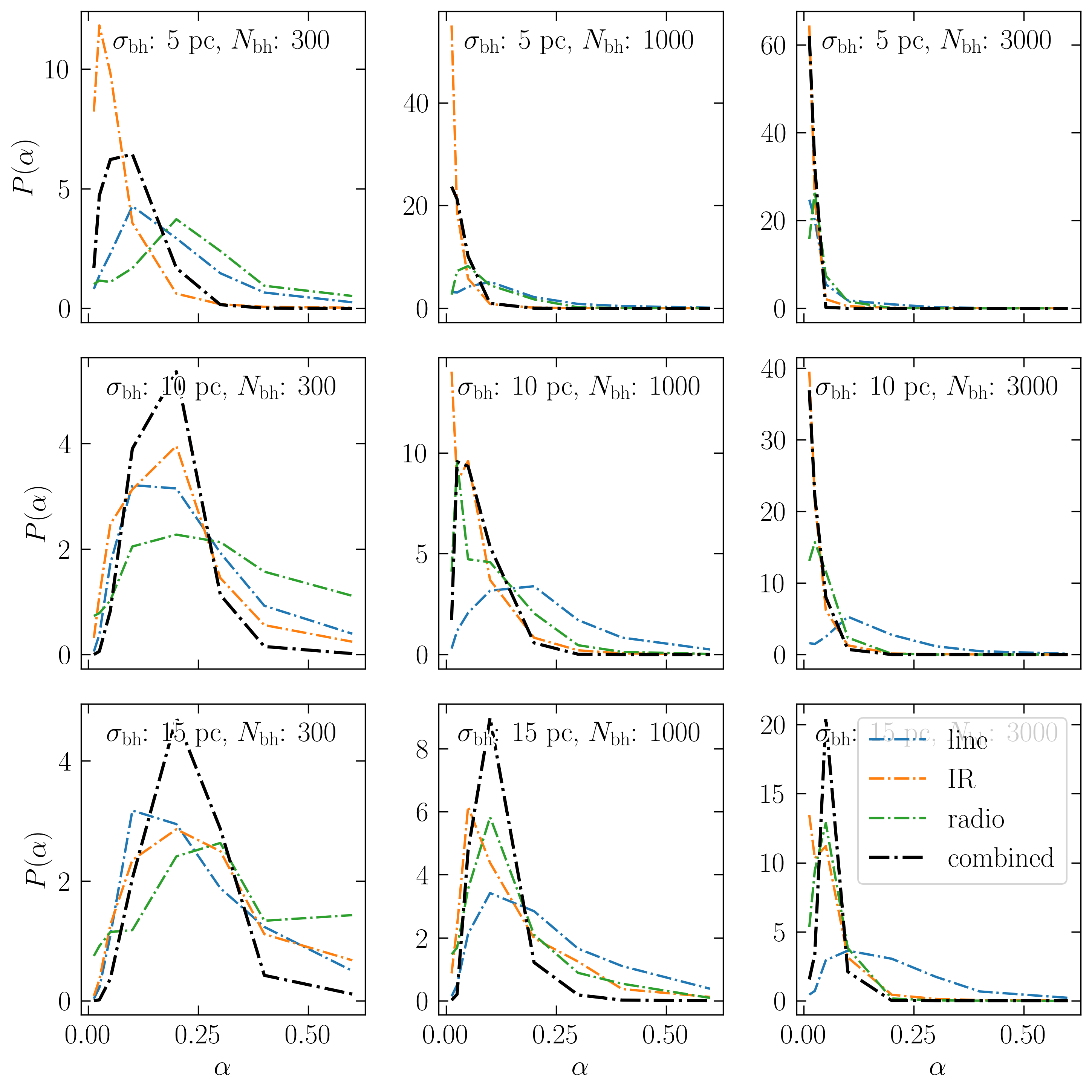}
        \caption{Same as Fig.~\ref{fig:alpha_microlensing}, but the error estimate on the magnification was set to $\sigma=0.14$ instead of $\sigma=0.24$.}
        \label{fig:alpha_microlensing_sigma_0p14}
\end{figure*}

        \label{sec:results:small_errors}
        We have already stressed the importance of accurate models in estimating the microlensing signal, as we use model magnifications as a no-microlensing baseline to estimate microlensing flux ratios. Throughout this work, we have allowed for a conservative 20\% error in the model flux ratios. Here we want to explore the different scenario in which we can fully trust the flux ratios predicted by the models, and therefore we have a smaller (14\%) error in the observed microlensing flux ratios, coming exclusively from errors in the observations. This also reflects a mixed scenario where model predictions are more reliable and/or measurement precision is higher than we have assumed. We highlight that such low errors in models are rather unrealistic \citep[cf.][]{Mediavilla:2024} and that this alternative scenario is explored mainly to show the impact of errors in the final outcome. We advise the reader to consider the previous analysis as the most realistic and likely scenario.

        The analysis involved repeating the microlensing signal estimation procedure under the assumption of this reduced error. The results, illustrated in Figure \ref{fig:alpha_microlensing_sigma_0p14}, indicate that a smaller error significantly increases the probability of detecting a non-zero fraction of mass in microlenses, denoted as $\alpha$. This can be understood by investigating Figures \ref{fig:magnification_maps_clustered_bh_histograms} and \ref{fig:cc_mag_histograms}. The radio region is, especially for small $N_\mathrm{bh}$ and large $\sigma_\mathrm{bh}$, large enough so that microlensing fluctuations get mostly `washed-out', meaning that the magnification histogram peaks sharply around $\alpha=0$. Any significant deviation of the microlensing magnification from 0 now automatically favors high fractions of mass $\alpha$; no matter whether this deviation is due to actual microlensing or due to modeling or measurement uncertainties.
        
\begin{figure}
    \centering
    \includegraphics[width=\linewidth]{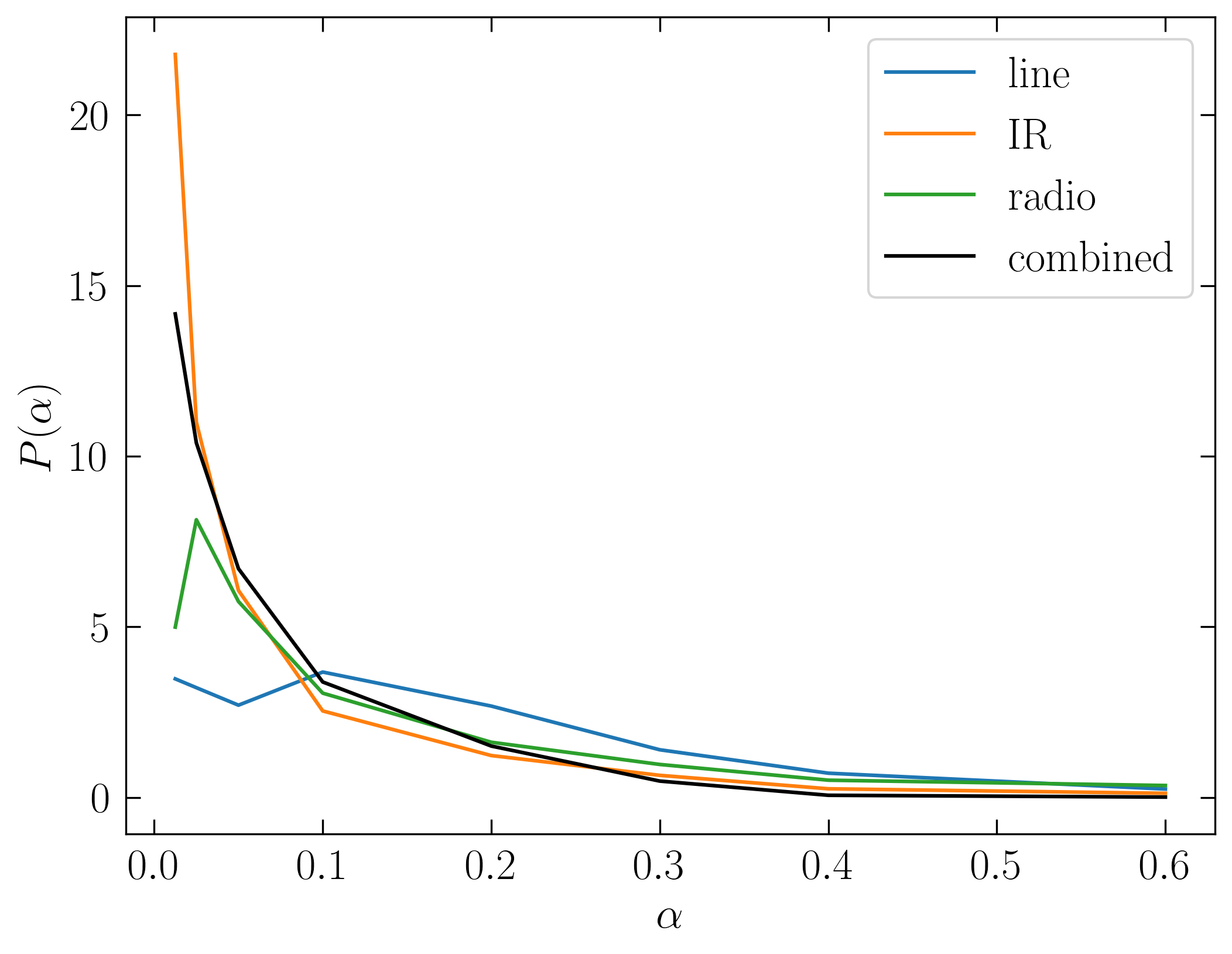}
    \caption{Same as Fig.~\ref{fig:frac_alpha_marginalized} assuming smaller errors for the anomalous flux ratios}
    \label{fig:frac_alpha_marginalized_nomeasurementerror}
\end{figure}

        Despite this increased probability, the permitted abundance of microlenses remains relatively low across most scenarios. Figure \ref{fig:frac_alpha_marginalized_nomeasurementerror} shows that only under the most favorable conditions, specifically with a low number of black holes ($N_{bh}=300$) and a large cluster size ($\sigma_{bh}=15$), does the microlensing mass fraction reach an upper limit of about 27\%. In most other configurations, the abundance is much lower, underscoring that even with unreasonably low errors, the presence of microlenses at significant levels is unlikely.
        
        These findings emphasize the importance of minimizing model errors to enhance the reliability of microlensing studies. By improving model accuracy, we can better distinguish between actual microlensing effects and those arising from observational or modeling uncertainties. This distinction is crucial for accurately estimating the fraction of mass in microlenses and understanding the underlying astrophysical processes.

        According to this investigation, we consider the possibility that model shortages or overestimation of their errors are reducing the contribution of PBHs to be very unlikely.
        
\section{Discussion}

Our simulations show that clustered BHs can act like giant pseudo-particles, enlarging the spatial scale of microlensing magnification patterns (i.e. "zooming in" the main features of microlensing magnification maps while the source size is kept fixed). This result 
extrapolates to the compact but finite clusters the consequences of the mass-length degeneracy of gravitational lensing, strictly applicable only to point particles \citep[see, e.g.,][]{2020ApJ...904..176E}. For extended sources, concentrating the lens distribution in a smaller number of more massive microlenses will increase the probability of high microlensing. Thus, considering more concentrated clusters than the ones analyzed in this work (either by increasing the number of BHs per cluster or by reducing the cluster size) will decrease the compatibility of a population of PBHs with observed data. This conclusion is independent of the mass of the individual BHs.

We have directly checked that the BLR, usually assumed to be insensitive to microlensing by single PBHs, could still experience strong microlensing under the action of these giant pseudo-particles constituted by the clustered PBHs. Consequently, we have been forced to introduce an alternative procedure by setting the non-microlensing baseline from the flux ratios predicted by lens modeling.
This alternative path presents the drawback of the large uncertainty in the estimate of the error of the flux-ratio predictions, and we have investigated its impact on our results. We have adopted a conservative estimate of $\sim$20\% for the lens model flux ratio error. A larger value would make the presence of clustered BHs even more unlikely. Although a smaller value is difficult to accept, it is interesting to explore the limiting case considering that models are very accurate, which we have done in Sect.~\ref{sec:results:small_errors}. 
This is equivalent to counting as a microlensing signal any possible error in the flux-ratio predictions by the models. 
While the results of this low error scenario do not rule out entirely the possible existence of a fraction of mass in clustered BHs, only in the most favorable configuration ($N_{bh}=300,\sigma_{bh}=15$), the permitted abundance presents an upper limit of 0.27 (0.42) at 68\% (95\%) confidence level, while it is much lower for most cases (see Fig.~\ref{fig:alpha_microlensing_sigma_0p14} in Sect.~\ref{sec:results:small_errors}).
Indeed, we can average over models to produce a single PDF similarly as we did before to produce Fig.~\ref{fig:frac_alpha_marginalized}. In this (unlikely) lower error scenario, the upper limit for the abundance of PBHs at 68\% (95\%) confidence level is 0.12 (0.28). We note that more concentrated clusters appear to permit only very small abundances of PBH. For more diluted clusters, the microlensing properties converge towards the ones of a uniform distribution of PBH. For these cases, one does not have to rely on a theoretical estimate for a no-magnification baseline. As previous studies exclude significant PBH abundances for uniform distributions, diluted PBH clusters do not appear to be a satisfactory dark matter model.

Leaving aside the very unlikely possibility of extremely accurate lens modeling, despite the large errors associated with flux-ratio predictions, the combination of probability distributions of 35 image-pairs confers a high statistical reliability to our main conclusions.

Thus, considering jointly the results of this work for clustered BHs with the ones from previous work for uniformly distributed BHs either based in optical or X-ray data \citep[cf.][]{2022ApJ...929..123E,2023ApJ...954..172E}, we can conclude that in the mass range from planetary, $\sim 0.001 M_\odot$, to intermediate, $\sim 100 M_\odot$, PBHs,  clustered or not, are very unlikely to be a main constituent of Dark Matter. 

\section{Conclusions }

We have studied the abundance of clustered PBHs of intermediate mass ($\sim$30$M_\odot$) according to the flux-ratio anomalies observed in the large-sized regions (insensitive to microlensing by stars) where the broad emission lines, the IR emission, and the radio emission are generated. We have compared the experimental flux-ratio anomalies (obtained using lens models to define the intrinsic flux-ratios) with the model predictions of microlensing caused by PBHs grouped in clusters of different numbers of PBHs and sizes. The main results of the present work are the following:
\begin{enumerate}

\item PBH clusters act like pseudo-particles of very large mass which give rise to the caustic structure that dominates the magnification maps. For clusters composed of a few hundred BHs, the Einstein radius considerably exceeds the size of the BLR, which can no longer be used to set the zero microlensing baseline.

\item The degree of concentration of the distribution of BHs in clusters/pseudo-particles (either increasing the number of BHs per cluster or reducing the size of the clusters) increases the likelihood of high microlensing magnifications. As the observed flux-ratio anomalies for these large sources (lines, IR, and radio emissions) are relatively small, this result disfavors the presence of clusters in lens galaxies. If these clusters were there, their microlensing effect would undoubtedly have been patent.

\item Our findings show that the likelihood of the presence of a population of massive PBHs increases with the progressive dissolution of the clusters. Anyhow, even in the most favorable case considered by us (apart from the investigations in Sect.~\ref{sec:results:small_errors} with unrealistic errors), we estimate a fraction of mass in PBHs of less than 7\% at 68\% confidence.

\item Although our main conclusion that clustered BHs can not significantly contribute to the Dark Matter is statistically very robust,
in Sect.~\ref{sec:results:small_errors} we have explored the implications of a systematic overestimate of model uncertainties. We confirm that our fiducial analysis is made with reasonable and well-motivated choices and that even considering (very unlikely) zero errors in the models, the predicted PBH cluster abundances cannot explain Dark Matter.

\item Combining the results from this work with previous investigations \citep{2022ApJ...929..123E,2017ApJ...836L..18M}, it appears that the $30\,\Msun$ mass-range, while very well-motivated by gravitational wave observations, is not a likely candidate for primordial black holes posing a significant fraction of dark matter.

\end{enumerate}

\begin{acknowledgements} 
We thank the anonymous referee for constructive comments that substantially improved the structure of this work. We are very grateful to Ch. Fassnacht, S. Vegetti, and J. Cohen for kindly providing us with the updated model parameters of B 0712+472. This research was supported by the grants PID2020-118687GB-C31, PID2020-118687GB-C32, and PID2020-118687GB-C33, financed by the Spanish Ministerio de Ciencia e Innovación through MCIN/AEI/10.13039/501100011033. J.J.V. is also
supported by project FQM-108
financed by Junta de Andalucía. HVA is supported by the grant PTA2021-020561-I, funded by MICIU/AEI/10.13039/501100011033 and by “ESF+”. SH acknowledges support from the German Research Foundation (DFG SCHN 342/13), the International Max-Planck Research School (IMPRS) and the German Academic Scholarship Foundation. SH is supported by the U.D Department of Energy, Office of Science, Office of High Energy Physics under Award Number DE-SC0019301. We acknowledge the use of the lux supercomputer at UC Santa Cruz, funded by NSF MRI grant AST 1828315. 
\end{acknowledgements}

\bibliographystyle{aa}
\bibliography{cite}
\appendix
\section{Notes on individual objects \label{appena}}
\label{app:indiv-systems}

\subsection{B 0128+437}
The source redshift for this system is $z_s$=3.124. There are two proposed lens redshifts ($z_l$) at 1.145 and 0.645; the former is deemed more probable \citep{Lagattuta:2010}, hence, it is the one we adopt for this study. Radio flux ratios have been obtained from \citet{2003ApJ...595..712K}. An anomaly in the flux ratio of image B in the radio wavelengths could be attributed to scatter-broadening within the interstellar medium of the lensing galaxy. However, super-resolution imaging at scales less than one milliarcsecond indicates potential substructure in the lens \citep{2004MNRAS.350..949B}.

\subsection{MG 0414+0534}
The source redshift of this system is $z_s$=2.639, with a lens redshift $z_l$=0.9584. Mid-infrared flux ratios are derived from \citet{Minezaki:2009,MacLeod:2013}, while radio flux ratios are drawn from \citet{Rumbaugh:2015}. A wavelength-dependent flux anomaly, previously suggesting the presence of a subhalo near image A2 \citep{MacLeod:2013}, appears to be clarified by the discovery of a dusty dark dwarf galaxy via ALMA by \citet{Inoue:2017}, exhibiting a redshift within the range of $0.5 < z < 1$. The model we utilize is based on \citet{MacLeod:2013}, which is additionally informed by the VLBI positions of four source components \citep{2000A&A...362..845R}.

\subsection{HE 0435–1223}
This system, with a source redshift $z_s = 1.689$ and a lens redshift $z_l = 0.4541$, presents a complex picture of gravitational lensing phenomena. Flux anomalies were reported in broad line flux ratios by \citet{Wisotzki:2003}, while radio flux ratios were found by \citet{Jackson:2015} to be consistent with a singular isothermal ellipsoid (SIE) model, provided an extended source ($\sigma \approx 300$ pc, although models with sources more than an order of magnitude smaller also fit reasonably well). Notably, these findings did not necessitate the presence of substructure in the lens.

Measurements by \citet{Nierenberg:2017} of [OIII] flux ratios concurred with those of \citet{Jackson:2015}. These measurements constrained potential NFW perturbers within approximately 1.0 (0.1) arcsec of the lensed images to a maximum mass within their central 600 pc of $10^8 (10^{7.2}) M_{600}/M_{\odot}$.

A comprehensive model that incorporated time delays between quasar images and the kinematics of the lens galaxy was provided by \citet{Wong:2017}. Flux ratios in the L' band were reported by \citet{Fadely:2012}.

We note that recent research presented in H0LiCOW Paper IV \citep{arxiv:1607.00382} constructed a more detailed lens model and required an additional external shear amplitude when including only G1 or all the galaxies G1-G5 in the model. However, we note that in order to achieve percent-level precision in Hubble constant measurements, their requirements for the lens model are a lot more strict than ours, so we do not adopt their lens model.

\subsection{RX J0911+0551}
The gravitational lensing system RX J0911+0551 has a source redshift $z_s=2.80$ and a lens redshift $z_l=0.769$. Radio flux ratios are taken from \citet{Jackson:2015}. Previous modeling of the system using a Singular Isothermal Ellipsoid (SIE) plus an external shear term ($\gamma$) and a Singular Isothermal Sphere (SIS) was conducted by \citet{Sluse:2012} and \citet{Schechter:2014}. However, \citet{Jackson:2015} reported a good fit using only an SIE and an external shear term, complemented by an extended source with a scale size of approximately 450 parsecs. We use here the model by \citet{Schechter:2014}.

\subsection{PG 1115+080}
The gravitational lensing system PG 1115+080 has a source redshift of $z_s= 1.722$ and a lens redshift of $z_l=0.3098$. Flux ratios of optical emission lines are derived from \citet{Popovic:2005}, while mid-infrared data for this system has been provided by \citet{Chiba:2005}. There are new radio measurements by \citet{2021MNRAS.508.4625H} which are not used in the present study because the measured size of the radio emission is much larger than the one used here.

\subsection{RXS J1131–1231}
For the gravitational lensing system RXS J1131–1231, the source redshift is $z_s=0.658$ and the lens redshift is $z_l=0.295$. [OIII] emission line flux ratios have been collected via both long-slit spectroscopy \citep{Sluse:2007} and Integral Field Spectroscopy (IFS) measurements \citep{Sugai:2007}. However, the long-slit spectroscopy measurements are influenced by the arc existing between images A and C. Detailed lens modeling that incorporates time delays and kinematics has been conducted by \citet{Suyu:2013}, \citet{Suyu:2014}, and \citet{Birrer:2016}. Additionally, a substructure quantification study for this system has been undertaken by \citet{Birrer:2017}. We use here the model by \citet{Schechter:2014}.

\subsection{B J1422+231}
The gravitational lensing system B J1422+231 is defined by a source redshift of $z_s=3.62$ and a lens redshift of $z_l=0.3366$. Radio flux ratios have been sourced from \citet{2003ApJ...595..712K}. Flux ratios of optical emission lines are available from \citet{Impey:1996}, while mid-infrared data has been procured from \citet{Chiba:2005}.

\subsection{B 1608+656}
The gravitational lensing system B 1608+656 presents a source redshift $z_s$=1.394 and a lens redshift $z_l$=0.6304. Radio flux ratios were obtained from \citet{Fassnacht:1999}. The lensing galaxy was modeled by \citet{Sluse:2012} as a combination of a Singular Isothermal Ellipsoid (SIE) and a Singular Isothermal Sphere (SIS), though the two components are likely interacting and are not strictly isothermal. \citet{Suyu:2009,Suyu:2010} employed more sophisticated lens modeling involving grid-based potential corrections of approximately 2\%, albeit at the cost of increased complexity in replication, so we stick with the model from \citet{Sluse:2012}.

\subsection{WFI J2033–4723}
The gravitational lensing system WFI J2033–4723 is characterized by a source redshift $z_s$=1.66 and a lens redshift $z_l$=0.66. Flux ratios of optical emission lines are taken from \citet{Morgan:2004}. \citet{2020MNRAS.498.1440R} presents a newer mass model by the H0LICOW collaboration, which we again disregard in favor of \citet{Schechter:2014} as we do not require that level of precision.

\subsection{Q 2237+0305}
The gravitational lensing system Q 2237+0305 has a source redshift $z_s$=1.695 and a lens redshift $z_l$=0.0395. [OIII] emission line flux ratios are derived from \citet{Wayth:2005}, while mid-infrared and model flux ratios have been provided by \citet{Vives-Arias:2016}. Radio flux ratios were obtained from \citet{Falco:1996}. A comprehensive model of the system, inclusive of the bulge, bar, disk, and halo components and employing kinematics as constraints, was developed by \citet{Trott:2010}.

\subsection{B 1555+375}
In the case of the gravitational lensing system B 1555+375, the models by \citet{Xu:2015} employ values of $z_s$ = 1.59 and $z_l$ = 0.6. Radio flux ratios for the system have been taken from \citet{2003ApJ...595..712K}. The inclusion of the edge-on disk of the lensing galaxy within the model successfully reproduces most of the observed radio flux ratio anomaly, as documented by \citet{Hsueh:2016}. We use this latter model in the present work.

\subsection{B 0712+472}
B 0712+472 is a gravitational lens system with a source redshift of $z_s=1.34$ and lens redshift of $z_l=0.406$ \citep{2003ApJ...595..712K}. The flux ratio anomaly in this system was studied in detail by \cite{Hsueh:2017}. They found that the anomaly can be explained by the presence of an edge-on disc in the lens galaxy, resolving the apparent discrepancy. In the present work, we use a model including an exponential disk kindly provided by J. Cohen (priv. comm.).

\end{document}